# Magnetic frustration in the sawtooth lattice of Zn$Ln_2$S$_4$ ($Ln$ = Er, Tm, Yb) olivines

G.C. Lau,[1] B.G. Ueland,[2] R.S. Freitas,[2] M. L. Dahlberg,[2] P. Schiffer,[2] and R.J. Cava[1]

[1]Department of Chemistry, Princeton University, Princeton, New Jersey 08544

[2]Department of Physics and Materials Research Institute, Pennsylvania State University, University Park, Pennsylvania 16802

**Abstract**

We report the magnetic properties of the Zn$Ln_2$S$_4$ ($Ln$ = Er, Tm, Yb) olivines, in which the magnetic lanthanide ions are in a potentially frustrated geometry consisting of sawtooth chains of corner-sharing triangles. Fits to the high temperature magnetic susceptibility yielded Curie-Weiss temperatures of $\theta_W \approx$ -4, -13, and -75 K for the Er, Tm, and Yb compounds respectively. Antiferromagnetic interactions indicated by these negative $\theta_W$ values, along with the absence of magnetic long range order above T = 1.8 K, suggest that geometrical frustration of magnetic ordering exists in these sulfide olivines as has been observed previously in the rare earth chalcogenide spinels.

PACS number(s): 75.50.-y, 61.10.Nz

**Introduction**

Geometrical magnetic frustration can occur in compounds where the magnetic lattice, often composed of triangles or tetrahedra, suppresses the long range ordering of spins and gives rise to many degenerate magnetic ground states. Magnetic atoms arranged in a corner-sharing tetrahedral motif are found in pyrochlore and spinel lattices, and studies have yielded many examples of novel physics associated with frustration in such structures.[1,2] Examples of such frustration have been found in the transition metal sulfide spinels,[3,4,5] with frustration of the structural degrees of freedom even leading to a ferroelectric state in $CdCr_2S_4$.[6] The rare earth spinels, $CdLn_2Se_4$ ($Ln$ = Dy, Ho) and $CdLn_2S_4$ ($Ln$ = Ho, Er, Tm, Yb), also appear to be frustrated, showing no magnetic ordering down to $T = 2$ K and only reaching partial saturation of the full expected moment up to applied fields of $H = 9$ T.[7] The work on these thio-spinels suggests that investigating the magnetic properties of sulfides in the olivine structure, a structure type closely related to the spinel structure, should be of interest from the viewpoint of magnetic frustration.

Considering only the B-site, $AB_2X_4$ spinels can be built up by alternating two different types of layers: kagomé nets and triangular layers. However, in the $AB_2X_4$ olivines, only one type of layer exists: an incomplete kagomé net resulting in sawtooth chains of alternating triangles. These triangles are nearly equilateral, creating a frustrated geometry. Figure 1 shows the magnetic lattice of an $AB_2X_4$ olivine with bond distances given for $ZnTm_2S_4$ as an example.[8] Two crystallographically distinct sites exist for the $Ln$ atoms, denoted as M1 and M2. The M1 sites sit along the shorter backbone segments of the chains. Each M1 site has 4 nearest neighbor spins that form the sawtooth configuration. The partial kagomé net can be revealed by connecting the sawtooth chains with a column of missing Ln atoms. Layers of isolated chains in the *ab* plane stack in an *ABAB* sequence along the *c* direction. The sulfur atoms have not been shown, but they form edge-sharing octahedra around the *Ln* cations.

Transition metal oxide olivines have been well characterized.[9,10,11,12,13,14] Both $Mn_2SiO_4$ and $Mn_2GeO_4$ were found to exhibit magnetic frustration with antiferromagnetic transition temperatures well below the Curie-Weiss temperature magnitude $|\theta_W|$. The frustration was found to be less pronounced in $Fe_2SiO_4$, and nonexistent in $Co_2SiO_4$. The noncollinear spin behavior may be attributed to the effect of competing exchange interactions or the strong coupling to the crystal field seen in many transition metal compounds. The inherent differences in the effects of crystal fields on lanthanide ions, and the potential for much stronger dipole-dipole interactions, suggests that the rare earth olivines may exhibit unusual magnetic states.

The Zn$Ln_2$S$_4$ (*Ln* = Er, Tm, Yb) system represents a good candidate for the study of lanthanide olivines because the rare earth atoms fully occupy the frustrating lattice sites with no cation mixing.[8] The specific synthesis and structure of these compounds is described in earlier papers,[8,14] but, to our knowledge, thorough studies of their magnetic properties do not exist. Here we report the magnetic susceptibility $\chi(T)$ and magnetization $M(H)$ of $ZnEr_2S_4$, $ZnTm_2S_4$, and $ZnYb_2S_4$ over a temperature range of $T$ = 350 K – 1.8 K and magnetic fields up to $H$ = 9 T. We also compare the data to those from the closely related lanthanide spinels and other Er containing materials.

**Experimental**

Polycrystalline samples of Zn$Ln_2$S$_4$ ($Ln$ = Er, Tm, Yb) were synthesized as previously reported.[14] Stoichiometric amounts of Er (99.9%), Tm (99.9%), and Yb (99.9%) metals were each reacted separately with S (precipitated purified) in sealed evacuated quartz tubes at 800°C for 2 days to form $Ln_2$S$_3$. $Ln_2$S$_3$ was fired with ZnS (99.99%) in a one to one molar ratio in a sealed evacuated quartz tube at 1000°C for 3-5 days with intermittent shaking of the tube to form the olivine product. ZnEr$_2$S$_4$ formed in the same manner, but required a heating temperature of 1200°C. Sample purity was confirmed through x-ray powder diffraction using CuKα radiation and a diffracted beam monochromator.

We measured the dc magnetic susceptibility with a SQUID magnetometer (Quantum Design MPMS) on cooling over T = 350 – 1.8 K in an applied field of $H$ = 0.01 T. The ACMS option for a Quantum Design PPMS cryostat was employed for dc magnetization measurements as a function of field using an extraction technique. Curie-Weiss fits to the dc susceptibility data were performed over $T$ = 50 - 350 K.

**Results and Discussion**

Figures 2, 3, and 4 show the magnetic susceptibilities of ZnEr$_2$S$_4$, ZnTm$_2$S$_4$, and ZnYb$_2$S$_4$, respectively, as $\chi^{-1}(T)$. The high temperature portion of each data set fits the Curie-Weiss law [$\frac{1}{\chi} = \frac{1}{C}(T - \theta_W)$, where $C$ is the Curie constant], yielding the $\theta_W$ values and effective moments shown in Table I. The experimentally determined moments are consistent with the expected values for the free $Ln^{3+}$ ions. Unlike the transition metal olivines, none of our rare earth olivines show signs of long range magnetic order above $T$ = 1.8 K. ZnYb$_2$S$_4$ has a much larger |$\theta_W$| than the other two and shows deviations from the Curie-Weiss fit below 150 K. This behavior has been seen in other Yb containing

compounds and is normally attributed to low-lying crystal field levels becoming depopulated in our temperature range.[15]

The susceptibilities of these lanthanide olivines are similar to their spinel counterparts.[7] As with Cd$Ln_2$S$_4$ ($Ln$ = Er, Tm, Yb), the negative values of $\theta_W$ in the olivines suggests antiferromagnetic spin-spin interactions and, if not for the frustrated geometry, the onset of magnetic ordering at $T \sim |\theta_W|$. However, we observe no such ordering transitions down to $T$ = 1.8 K, although Figure 3 shows that $\chi^{-1}(T)$ of ZnTm$_2$S$_4$ displays a decreasing slope in $M(T)$ at the lowest temperatures, similar to what is observed in CdTm$_2$S$_4$.[7] The insets in Figures 2-4 illustrate little difference between the field-cooled and zero-field-cooled magnetizations. The lack of a bifurcation between the two measurements appears to exclude the possibility of a spin glass state or other frozen short-range ordered state in these materials above $T$ = 1.8 K. Coupled with the absence of long range ordering down to temperatures below $|\theta_W|$, the data suggests that geometric frustration is likely responsible for the lack of ordering of the $Ln$ spins.

The field dependent magnetizations at $T$ = 2 K of the three $Ln$ olivines are shown in Figure 5. These measurements were used to probe for unusual field induced states as, for example, have been seen in spin ice compounds.[16] In addition, the transition metal silicate olivines are known to possess different magnetic states at very high fields.[17] While the low $T$, high $H$ data taken on our materials does not illustrate any of these transitions, the three samples also did not achieve their full saturated values. The ZnEr$_2$S$_4$ data in particular display a clear magnetization plateau similar to what is seen in the spin ice pyrochlores and CdEr$_2$S$_4$.[18] This incomplete saturation may be a result of single-ion anisotropy in these randomly oriented polycrystalline samples.

Partial saturation in both ZnEr$_2$S$_4$ and CdEr$_2$S$_4$ suggested that other Er-containing materials might have previously been shown to exhibit the same magnetization plateau. Figure 6 shows *M*(*H*) data for a variety of polycrystalline Er containing compounds, all with incomplete saturations. Magnetization measurements on a dysprosium niobate pyrochlore have indicated that the average component of the moment along an applied field in a polycrystalline sample with local <111> Ising anisotropy is 0.5*J*, where *J* is the average exchange interaction between magnetic atoms.[19] Similarly, in polycrystalline samples with local planar anisotropy, the average component of the moment along the field is approximately 0.785*J*.[19] This implies that samples giving magnetization plateaus at 50% of the expected moment have Ising-like spins, while plateaus at closer to 80% saturation represent spins restricted to a plane. Upon comparison, our data suggest that CdEr$_2$S$_4$ has Ising-like spins, while ZnEr$_2$S$_4$ shows spins closer to planar anisotropy. Thus, we surmise the *M*(*H*) plateau in these Er compounds is likely the result of single-ion crystal field effects restricting the spin directions, creating Ising or planar anisotropies. However, orientation dependent studies of single crystals or highly diluted samples would be necessary to conclude if we have truly observed single-ion rather than collective behavior.

In summary, we report the magnetic properties of three lanthanide sulfide olivines, Zn*Ln*$_2$S$_4$ (*Ln* = Er, Tm, Yb). The absence of an observed transition down to temperatures well below |$\theta_W$| suggests that geometrical frustration in the sawtooth triangular chains of these compounds suppresses long-range magnetic order. Furthermore, our magnetization data may indicate the presence of planar anisotropy in ZnEr$_2$S$_4$. Data on these samples at lower temperatures and on single crystals would be of interest, since such data might reveal exotic ground states in analogy to those seen in geometrically frustrated oxides.

**Acknowledgement**

This research was supported by the National Science Foundation, under Grant No. DMR-035610. R.S.F. thanks the CNPq-Brazil for sponsorship. The authors gratefully acknowledge discussions with Shivaji Sondhi and Joseph Bhaseen.

| Table I. Weiss constants and magnetic moments determined from the Curie-Weiss fits of high temperature portions of the magnetic susceptibilities. | | | |
|---|---|---|---|
| Compound | $\theta_W$ (K) | $p$ Exptl. | $p$ Calc.$(g[J(J+1)]^{1/2})$ |
| $ZnEr_2S_4$ | $-3.6 \pm 0.3$ | $9.15 \pm 0.01$ | 9.59 |
| $ZnTm_2S_4$ | $-12.8 \pm 0.2$ | $7.19 \pm 0.01$ | 7.57 |
| $ZnYb_2S_4$ | $-75.2 \pm 0.9$ | $4.72 \pm 0.01$ | 4.54 |

**Figures**

**Fig. 1** The magnetic lattice of $ZnTm_2S_4$ with only Tm atoms shown. The M1 sites compose the backbone linear chain and connect to M2 sites to form alternating sawtooth isosceles triangles. The view from above of the *ab* plane (a) shows the isolated chains of solid black atoms connected by solid lines. Dashed lines illustrate the partial kagomé net by completing the slightly distorted lattice with a column of nonexistent atoms between the chains. Layers of these *ab* planes viewed from the side (b) show *ABAB* stacking along the *c* axis. Distances are given in Å.

**Fig. 2** (Color online) The inverse dc magnetic susceptibility versus temperature of $ZnEr_2S_4$ as measured in an applied field of $H = 100$ Oe. The solid line was obtained through a fit to the high temperature points. **Inset**: Field-cooled and zero-field-cooled susceptibility plotted versus temperature for $ZnEr_2S_4$.

**Fig. 3** (Color online) The inverse dc magnetic susceptibility versus temperature of $ZnTm_2S_4$ as measured in an applied field of $H = 100$ Oe. The solid line was obtained through a fit to the high temperature points. **Inset**: Field-cooled and zero-field-cooled susceptibility plotted versus temperature for $ZnTm_2S_4$.

**Fig. 4** (Color online) The inverse dc magnetic susceptibility versus temperature of ZnYb$_2$S$_4$ as measured in an applied field of $H$ = 100 Oe. The solid line was obtained through a fit to the high temperature points. **Inset**: Field-cooled and zero-field-cooled susceptibility plotted versus temperature for ZnYb$_2$S$_4$.

**Fig 5** (Color online) The normalized magnetizations versus applied field for the three olivines studied at $T$ = 2 K up to applied fields of $H$ = 9 T. The magnetizations have been normalized by the values expected for fully saturated free $Ln^{3+}$ ions ($M = gJ$).

**Fig 6** (Color online) The normalized magnetizations versus applied field for Er containing materials studied at $T$ = 2 K up to applied fields of $H$ = 9 T. The magnetizations have been normalized by the values expected for fully saturated free Er$^{3+}$ ions ($M = gJ$).

**Figure 1**

(a)

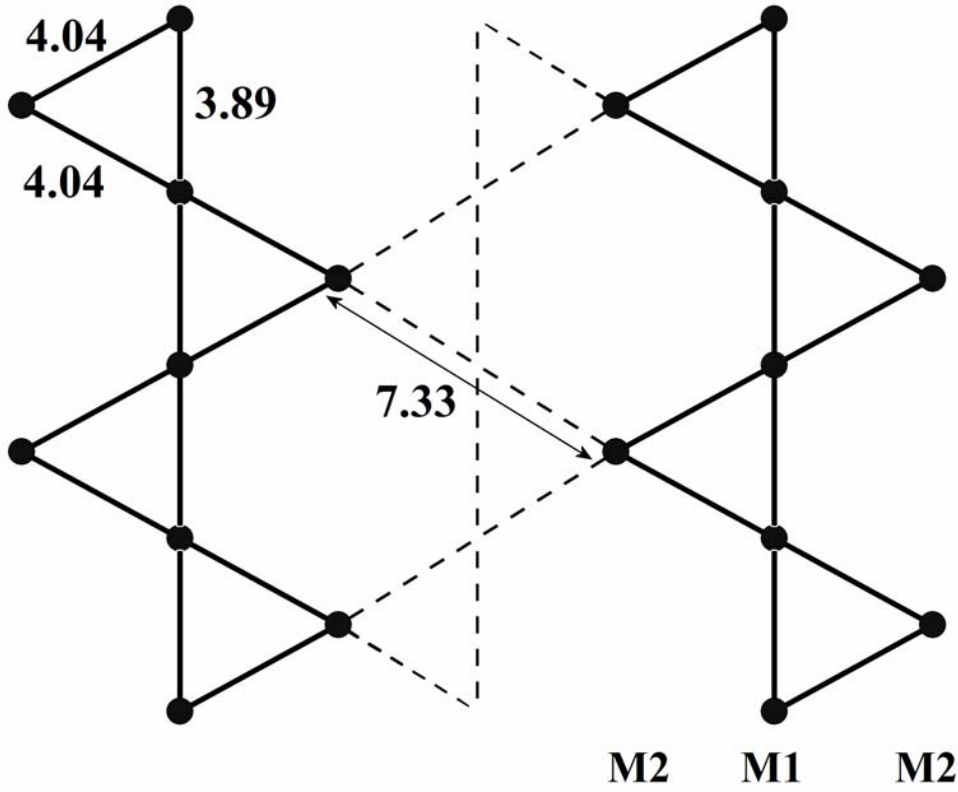

(b)

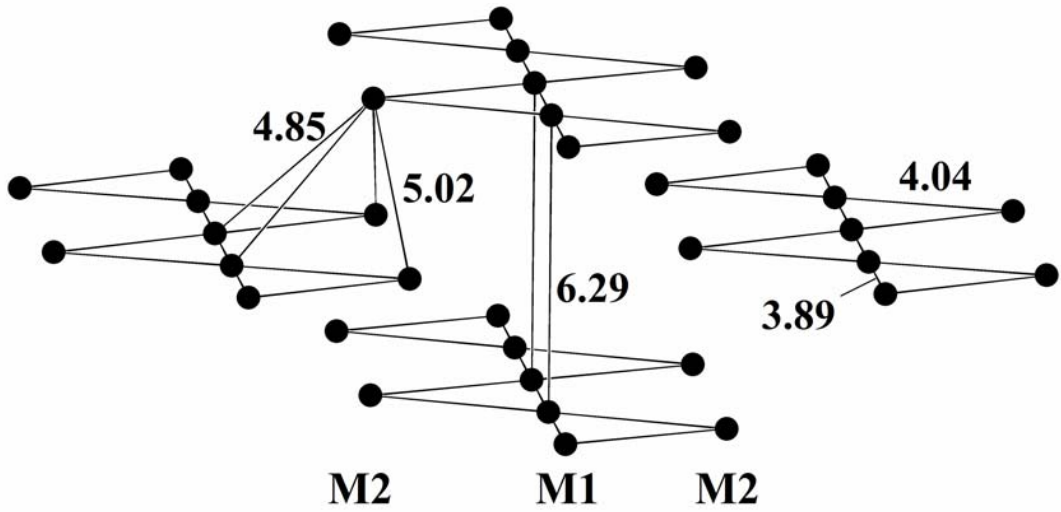

Figure 2

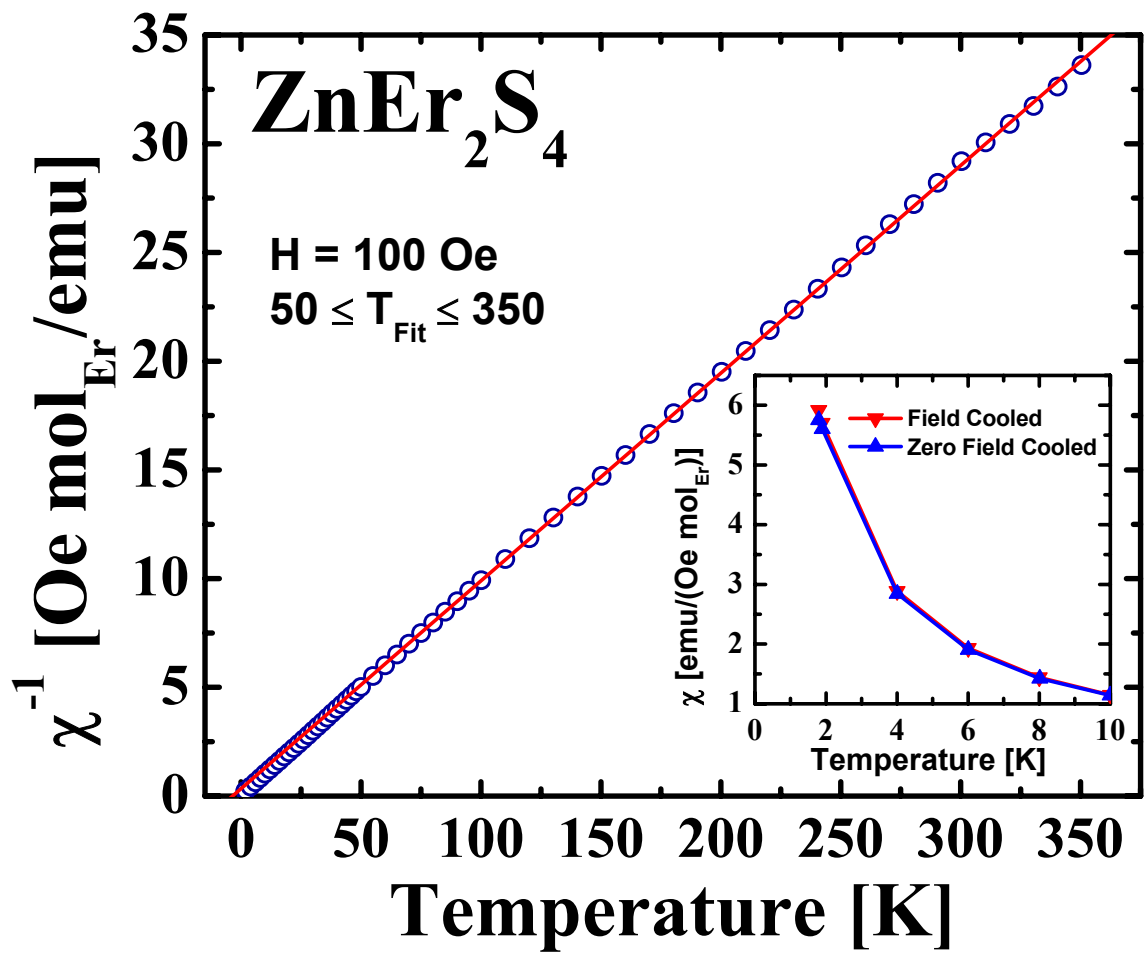

Figure 3

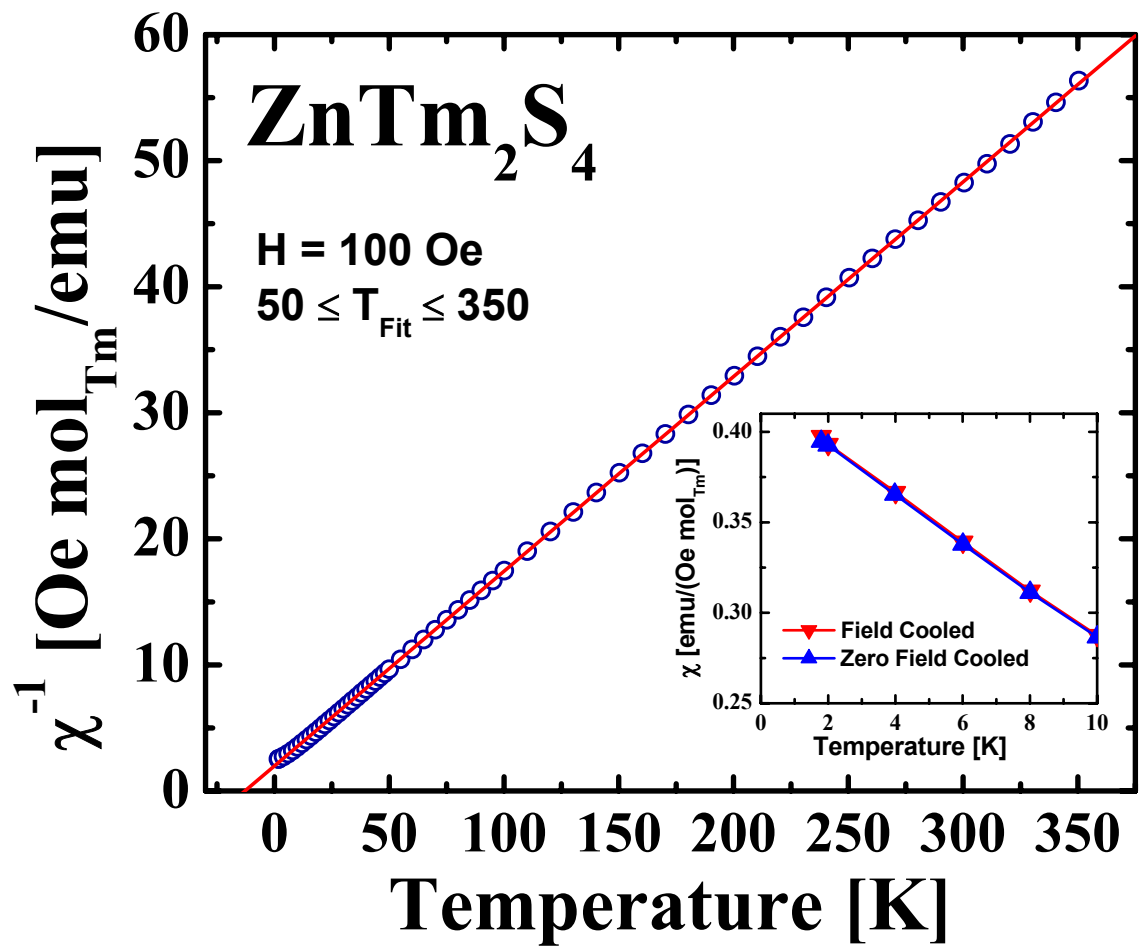

**Figure 4**

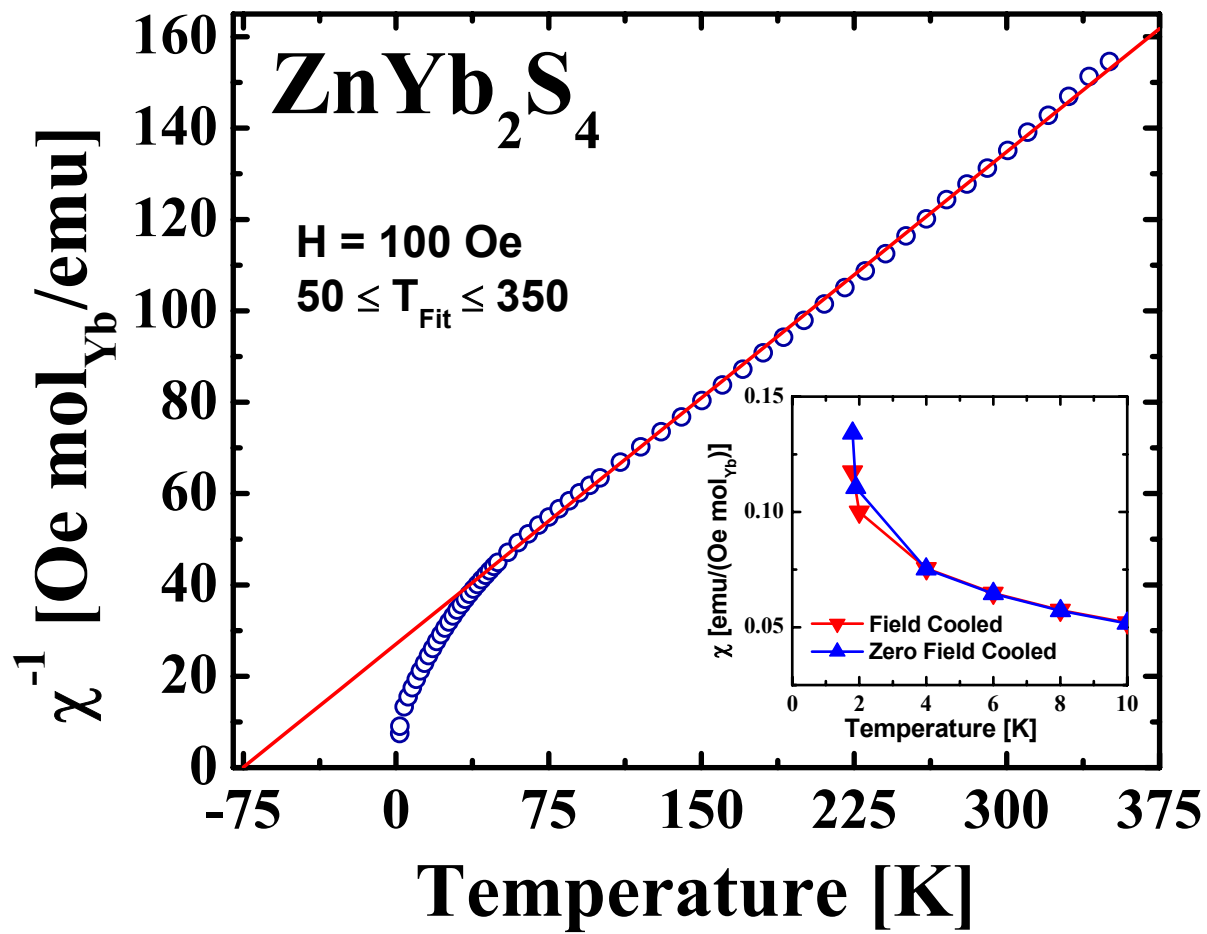

Figure 5

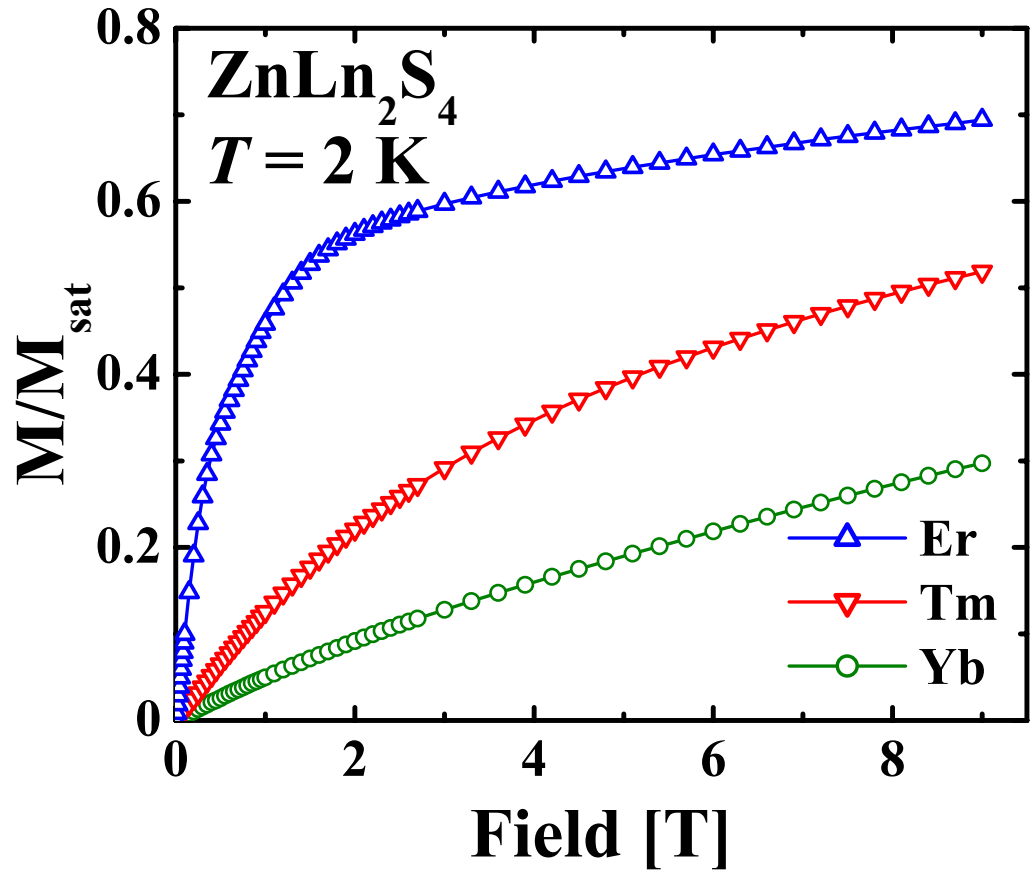

**Figure 6**

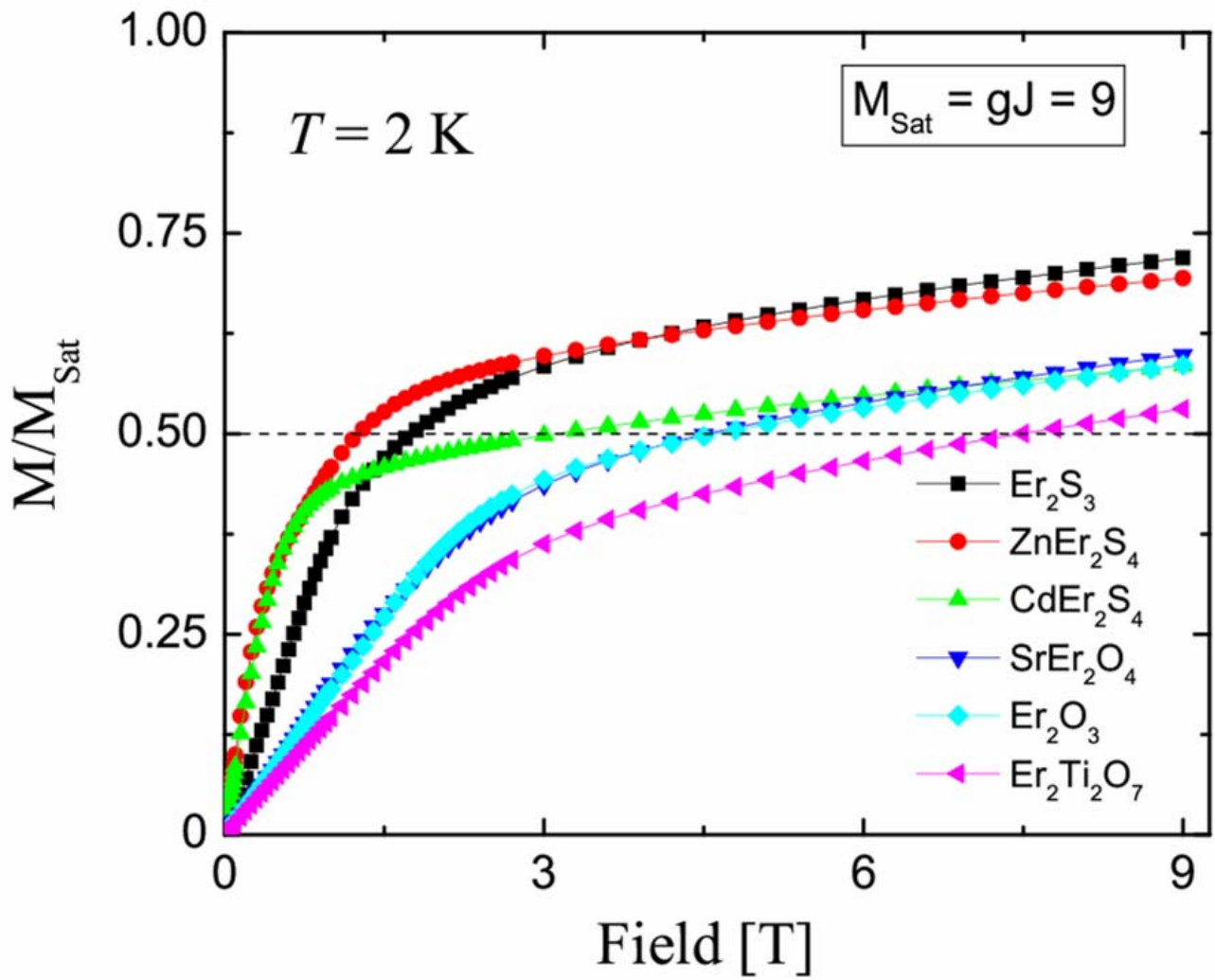